\documentstyle[times, psfig]{article}
\newcommand{\ls}[1]
   {\dimen0=\fontdimen6\the\font 
    \lineskip=#1\dimen0
    \advance\lineskip.5\fontdimen5\the\font
    \advance\lineskip-\dimen0
    \lineskiplimit=.9\lineskip
    \baselineskip=\lineskip
    \advance\baselineskip\dimen0
    \normallineskip\lineskip
    \normallineskiplimit\lineskiplimit
    \normalbaselineskip\baselineskip
    \ignorespaces
   }

\newcommand {\ix} {\hspace*{2em}}

\ls{1}

\begin{document}

\begin{titlepage}
\title{Optimizing Noncontiguous Accesses in MPI-IO\thanks{This work
was supported by the Mathematical, Information, and 
Computational Sciences Division subprogram of the Office of
Advanced Scientific Computing Research, U.S.\ Department of Energy,
under Contract W-31-109-Eng-38; and by the Scalable I/O Initiative, a
multiagency project funded by the Defense Advanced Research Projects
Agency (contract number DABT63-94-C-0049), the Department of Energy,
the National Aeronautics and Space Administration, and the National
Science Foundation. We thank the Center for Advanced Computing Research
at California Institute of Technology, the National Center for
Supercomputing Applications at the University of Illinois, and the
National Aerospace Laboratory (NLR) in Holland for providing access to
their machines.}}

\author{{\em Rajeev Thakur} \ix {\em William Gropp} \ix {\em Ewing Lusk} \\
Mathematics and Computer Science Division \\
Argonne National Laboratory \\
9700 S. Cass Avenue \\
Argonne, IL 60439, USA \\
\\
{\bf Corresponding Author}: William Gropp \\
Tel: (630) 252-4318, Fax: (630) 252-5986 \\
email: {\tt gropp@mcs.anl.gov}
}

\date{}
\maketitle

\begin{abstract}
The I/O access patterns of many parallel applications consist of
accesses to a large number of small, noncontiguous pieces of data. If
an application's I/O needs are met by making many small, distinct I/O
requests, however, the I/O performance degrades drastically. To avoid
this problem, MPI-IO allows users to access noncontiguous data with a
single I/O function call, unlike in Unix I/O. In this paper, we
explain how critical this feature of MPI-IO is for high performance
and how it enables implementations to perform optimizations.  We first
provide a classification of the different ways of expressing an
application's I/O needs in MPI-IO---we classify them into four {\em
levels}, called level~0 through level~3.  We demonstrate that, for
applications with noncontiguous access patterns, the I/O performance
improves dramatically if users write their applications to make
level-3 requests (noncontiguous, collective) rather than level-0
requests (Unix style). We then describe how our MPI-IO implementation,
ROMIO, delivers high performance for noncontiguous requests. We
explain in detail the two key optimizations ROMIO performs: data
sieving for noncontiguous requests from one process and collective I/O
for noncontiguous requests from multiple processes.  We describe how
we have implemented these optimizations portably on multiple machines
and file systems, controlled their memory requirements, and also achieved
high performance. We demonstrate the performance and portability with
performance results for three applications---an
astrophysics-application template (DIST3D), the NAS BTIO benchmark,
and an unstructured code (UNSTRUC)---on five different parallel
machines: HP Exemplar, IBM SP, Intel Paragon, NEC SX-4, and SGI
Origin2000.
\end{abstract}

\thispagestyle{empty}

\end{titlepage}

\section{Introduction}
I/O is a major bottleneck in many parallel applications. Although the
I/O subsystems of parallel machines may be designed for high
performance, a large number of applications achieve only about a tenth
or less of the peak I/O bandwidth. One of the main reasons for poor
application-level I/O performance is that I/O systems are optimized
for large accesses (on the order of megabytes), whereas many parallel
applications make lots of small requests (on the order of kilobytes or
even less). These small requests occur for the following reasons:
\begin{itemize}
\item In many parallel applications (for example, those that access
distributed arrays from files) each process needs to access a large
number of relatively small pieces of data that are not contiguously
located in the 
file~\cite{bayl96a,cran95a,nieu96c,smir96a,smir98a,thak96c}.
\item Most parallel file systems have a Unix-like API (application
programming interface) that 
allows a user to access only a single, contiguous chunk of data at a
time from a file.\footnote{Unix does have functions {\tt readv} and
{\tt writev}, but they allow noncontiguity only in memory and not in
the file. POSIX has a function {\tt lio\_listio} that allows the user
to specify a list of requests at a time. However, the requests in the
list can be a mixture of reads and writes, and the POSIX standard
says that each of the requests will be submitted as a
separate nonblocking request~\cite{posi96a}. Therefore, POSIX
implementations cannot optimize I/O for the entire list of requests, for
example, by performing data sieving as described in 
Section~\ref{sec:siev}. Furthermore, since the 
{\tt lio\_listio} interface is not collective, implementations cannot
perform collective I/O.}  Noncontiguous data sets must therefore be
accessed by making separate function calls to access each individual
contiguous piece.
\end{itemize}

With such an interface, the file system cannot easily detect the
overall access pattern of one process individually or that of a group
of processes collectively. Consequently, the file system is
constrained in the optimizations it can perform. Many parallel file
systems also provide their own extensions to or variations of the
traditional Unix interface, and these variations make programs
nonportable.

To overcome the performance and portability limitations of existing
parallel I/O interfaces, the MPI Forum (made up of parallel-computer
vendors, researchers, and applications scientists) defined a new
interface for parallel I/O as part of the MPI-2
standard~\cite{mpi97a}. This interface is commonly referred to as
MPI-IO. MPI-IO is a rich interface with many features designed
specifically for performance and portability. Multiple implementations
of MPI-IO, both portable and machine specific, are
available~\cite{fine96a,jone96a,pros96a,sand96a,thak97a}. To avoid the
above-mentioned problem of many distinct, small I/O requests, MPI-IO
allows users to specify the entire noncontiguous access pattern and
read or write all the data with a single function call. MPI-IO also
allows users to specify collectively the I/O requests of a group of
processes, thereby providing the implementation with even greater
access information and greater scope for optimization.

A simple way to port a Unix I/O program to MPI-IO is to replace all
Unix I/O functions with their MPI-IO equivalents. For applications
with noncontiguous access patterns, however, such a simple port is unlikely
to improve performance. In this paper, we demonstrate that to get significant
performance benefits with MPI-IO, users must use some of MPI-IO's
advanced features, particularly noncontiguous accesses and collective
I/O.

An application can be written in many different ways with MPI-IO.  We
provide a classification of the different ways of expressing an
application's I/O access pattern in MPI-IO. We classify them into four
{\em levels}, called level~0 through level~3. We explain why, for high
performance, users should write their application programs to make
level-3 MPI-IO requests (noncontiguous, collective) rather than
level-0 requests (Unix style). Similarly, I/O libraries, such as
HDF5~\cite{hdf-www}, that are written on top of MPI-IO should also
strive to make level-3 requests.

We describe how our portable implementation of MPI-IO, called ROMIO,
delivers high performance when the user makes noncontiguous,
collective I/O requests.  We explain in detail the two key
optimizations ROMIO performs: data sieving for noncontiguous requests
from one process and collective I/O for noncontiguous requests from
multiple processes. We describe how we have implemented these
optimizations portably on multiple machines and file systems, controlled
their memory requirements, and also achieved high performance. We
demonstrate the performance and portability with performance results
for three applications on five different parallel machines: HP
Exemplar, IBM SP, Intel Paragon, NEC SX-4, and SGI Origin2000.  The
applications we used are the following:
\begin{enumerate}
\item DIST3D, a template representing the I/O access pattern in an
astrophysics application, ASTRO3D~\cite{thak96c}, from the University
of Chicago. This application does a large amount of I/O and is
representative of applications that access distributed arrays.
\item The NAS BTIO benchmark~\cite{fine96a}, a well-known MPI-IO benchmark 
developed at NASA Ames Research Center.
\item An unstructured code (UNSTRUC) from Sandia National
Laboratories that is representative of applications that have 
irregular access patterns.
\end{enumerate}

The rest of this paper is organized as follows. In
Section~\ref{sec:noncontig}, we explain how MPI-IO supports
noncontiguous file accesses. In Section~\ref{sec:class}, we present a
classification of the different ways of expressing an application's
I/O access pattern in MPI-IO. We describe our MPI-IO implementation,
ROMIO, in Section~\ref{sec:impl}. In Sections~\ref{sec:siev}
and~\ref{sec:coll}, we describe in detail how data sieving and
collective I/O are implemented in ROMIO. Performance results are
presented and analyzed in Section~\ref{sec:perf}, followed by conclusions in
Section~\ref{sec:conc}.

\section{Noncontiguous Accesses in MPI-IO\label{sec:noncontig}}
In MPI, the amount of data a function sends, receives, reads, or
writes is specified 
in terms of instances of a {\em datatype}~\cite{mpi95a}.  Datatypes in
MPI are of two kinds: basic and derived. Basic datatypes 
correspond to the basic datatypes in the host programming
language---integers, floating-point numbers, and so forth. In
addition, MPI provides datatype-constructor functions to create
derived datatypes consisting of multiple basic datatypes located
either contiguously or noncontiguously. The
different kinds of datatype constructors in MPI are as follows:
\begin{itemize}
\item {\bf contiguous} Creates a new datatype consisting of
contiguous copies of an existing datatype.
\item {\bf vector/hvector} Creates a new datatype consisting of
equally spaced copies of existing datatype.
\item {\bf indexed/hindexed/indexed\_block} Allows replication of a
datatype into a sequence of blocks, each containing multiple copies of
an existing datatype; the blocks may be unequally spaced.
\item {\bf struct} The most general datatype constructor, which
allows each block to consist of replications of different datatypes.
\item {\bf subarray} Creates a datatype that corresponds to a
subarray of a multidimensional array.
\item {\bf darray} Creates a datatype that describes a process's local
array obtained from a regular distribution a multidimensional global array.
\end{itemize}
The datatype created by a
datatype constructor can be used as an input datatype to another
datatype constructor. Any noncontiguous data layout can therefore be
represented in terms of a derived datatype.

MPI-IO uses MPI datatypes to describe the data layout in the user's
buffer in memory and also to define the data layout in the file. The
data layout in memory is specified by the {\tt datatype} argument in
each read/write function in MPI-IO. The data layout in the file is
defined by the {\em file view}. When the file is first opened, the
default file view is the entire file; that is, the entire file is
visible to the process, and data will be read/written contiguously
starting from the location specified by the read/write function.  A
process can change its file view at any time by using the function
{\tt MPI\_File\_set\_view}, which takes as argument an MPI datatype
called the {\em filetype}. From then on, data will be read from or
written to only 
those parts of the file specified by the filetype; any ``holes'' will
be skipped. The file view and the data layout in memory can be defined
by using any MPI basic or derived datatype; therefore, any general
noncontiguous access pattern can be compactly represented.

Several studies have shown that, in many parallel applications, each
process needs to access a number of relatively small, noncontiguous
portions of a
file~\cite{bayl96a,cran95a,nieu96c,smir96a,thak96c}. From a
performance perspective, it is critical that the I/O interface can
express such an access pattern, as it enables the implementation to
optimize the I/O request. The optimizations typically allow the
physical I/O to take place in large, contiguous chunks, even though
the user's request may be noncontiguous. MPI-IO's file views,
therefore, are critical for performance. Users must ensure that they
describe noncontiguous access patterns in terms of a file view and
then call a single I/O function; they must not try to access each
contiguous portion separately as in Unix I/O.

\section{A Classification of I/O Request Structures \label{sec:class}}
Any application has a particular ``I/O access pattern'' based on
its I/O needs. The same I/O access pattern can be presented
to the I/O system in different ways, however, depending on which I/O functions the
application uses and how. We classify the different ways of expressing
I/O access patterns in MPI-IO into four ``levels,'' level~0 through level~3.
We explain this classification with the help of an example, accessing
a distributed array from a file, which is a common access pattern in
parallel applications. (One of the benchmark applications we used for
performance evaluation, DIST3D, has such an access pattern.)

\begin{figure}
\centerline{\psfig{figure=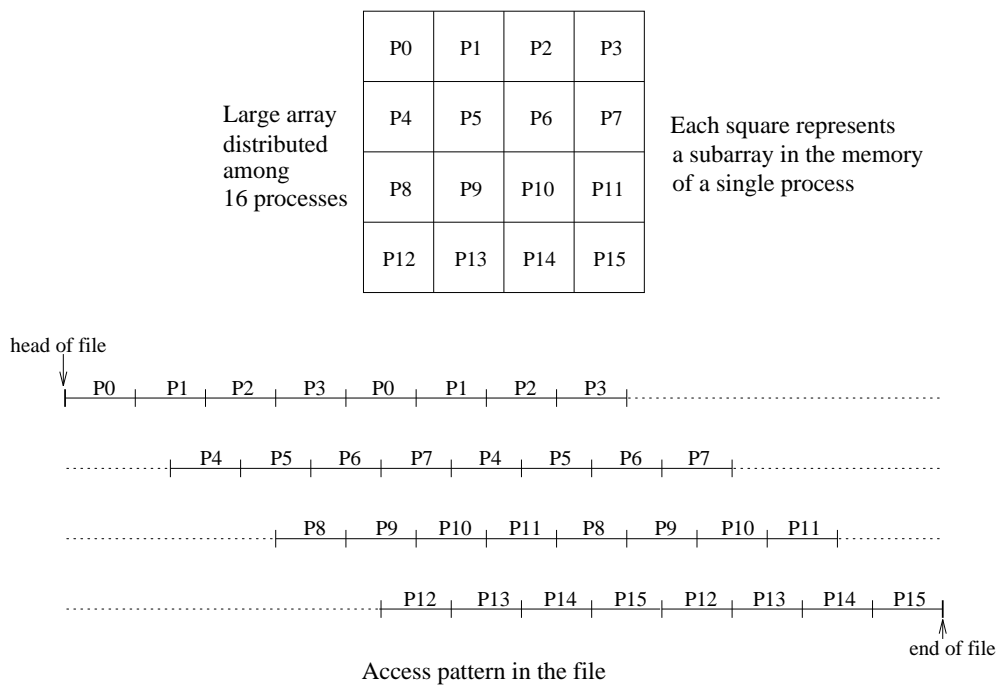,height=3.5in}}
\caption{Distributed-array access}
\label{fig:access}
\end{figure}

Consider a two-dimensional array distributed among 16~processes in a
(block, block) fashion as shown in Figure~\ref{fig:access}.  The array
is stored in a single file corresponding to the global array in
row-major order, and each process needs to read its local array from
the file. (Note that the file could be physically distributed among
disks, but appears to the program as a single logical file.) The data
distribution among processes and the array storage order in the file
are such that the file contains the first row of the local array of
process~0, followed by the first row of the local array of process~1,
the first row of the local array of process~2, the first row of the
local array of process~3, then the second row of the local array of
process~0, the second row of the local array of process~1, and so
on. In other words, the local array of each process is located
noncontiguously in the file.

\begin{figure}
\centerline{\psfig{figure=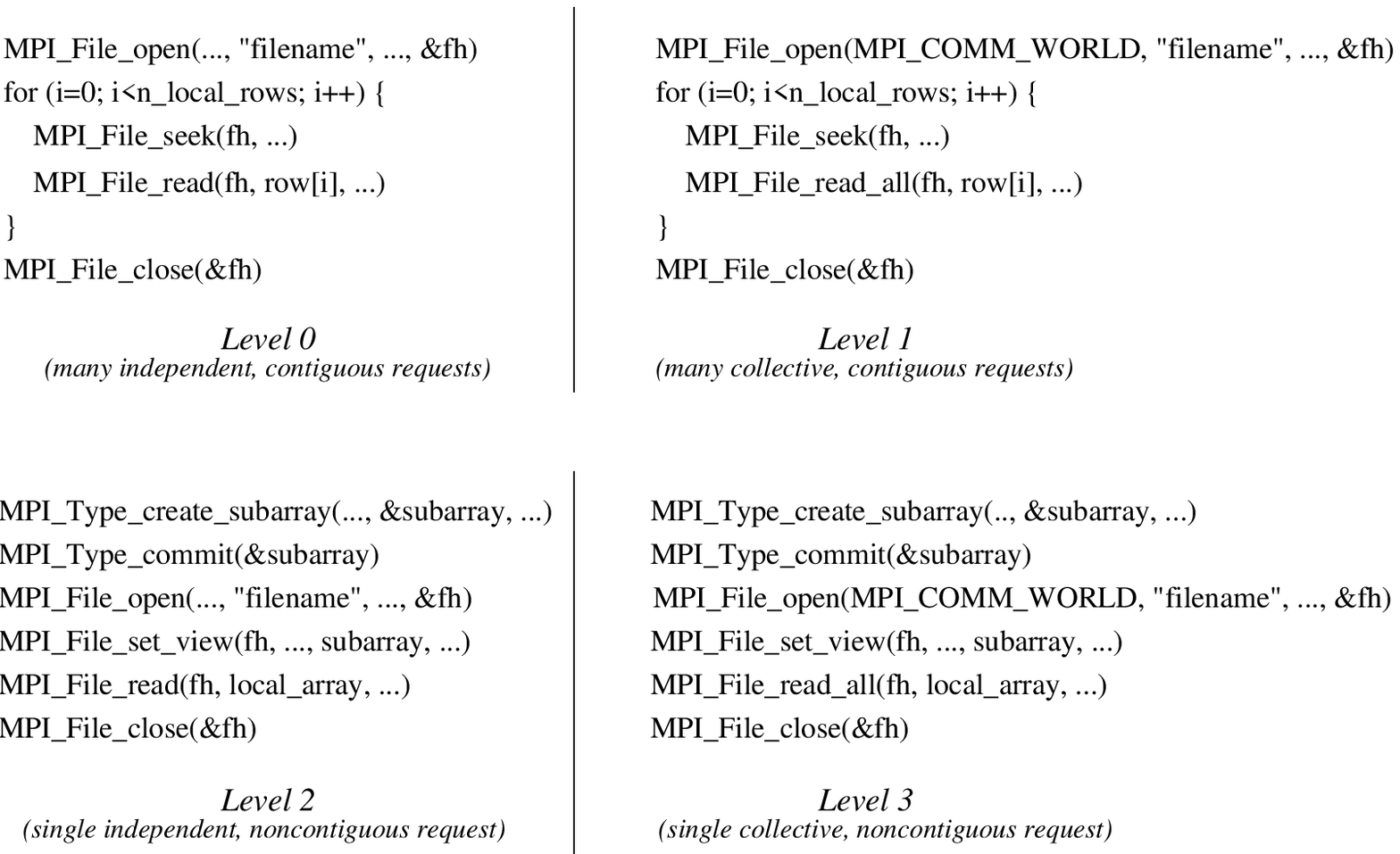,width=5.75in}}
\caption{Pseudo-code that shows four ways of accessing the data in
Figure~\protect\ref{fig:access} with MPI-IO}
\label{fig:code}
\end{figure}

Figure~\ref{fig:code} shows four ways in which a user can write a
program in which each process reads its local array from this file
using MPI-IO. In level~0, each process does Unix-style
accesses---one independent read request for each row in the local
array. Level~1 is similar to level~0 except that it uses
collective I/O functions, which indicates to the implementation that
all processes that together opened the file will call this function,
each with its own access information.  Independent I/O functions, on
the other hand, convey no information about what other processes will
do. In level~2, each process creates an MPI derived datatype to describe
the noncontiguous access pattern, defines a file view, and calls
independent I/O functions. Level~3 is similar to level~2 except that
it uses collective I/O functions.

\begin{figure}
\centerline{\psfig{figure=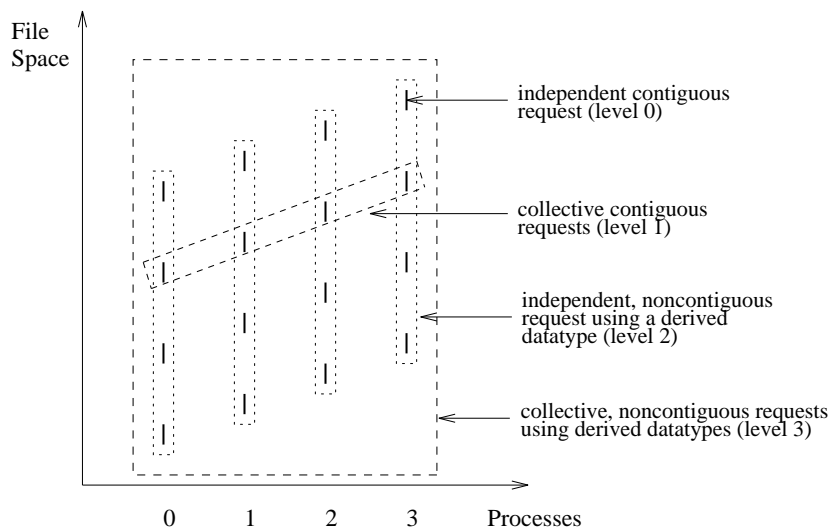,height=2.7in}}
\caption{The four levels representing increasing amounts of data per request}
\label{fig:graph}
\end{figure}

The four levels represent increasing amounts of data per request,
as illustrated in Figure~\ref{fig:graph}.\footnote{In this figure,
levels~1 and~2 represent the same amount of data per request, but, in
general, when the number of noncontiguous accesses per process is
greater than the number of processes, level~2 represents more data
than level~1.} The more the amount of data per request, the greater is
the opportunity for the implementation to deliver higher
performance. Users should therefore strive to express their I/O
requests as level~3 rather than level~0. How good the performance is
at each level depends, of course, on how well the implementation takes
advantage of the extra access information at each level.

If an application needs to access only large, contiguous pieces of
data, level~0 is equivalent to level~2, and level~1 is equivalent to
level~3. Users need not create derived datatypes in such cases, as
level-0 requests themselves will likely perform well.  Most parallel
applications, however, do not fall into this category. Several studies
of I/O access patterns in parallel
applications~\cite{bayl96a,cran95a,nieu96c,smir96a,smir98a,thak96c}
have shown that each process in a parallel program may need to access
a number of relatively small, noncontiguous portions of a file. From a
performance perspective, it is critical that the I/O interface can
express such an access pattern, as it enables the implementation to
optimize the I/O request. The optimizations typically allow the
physical I/O to take place in large, contiguous chunks, even though
the user's request may be noncontiguous. Users, therefore, should
ensure that they describe noncontiguous access patterns in terms of a
file view and then call a single I/O function; they should not try to
access each contiguous portion separately as in Unix
I/O. Figure~\ref{fig:example} shows the detailed code for creating a
derived datatype, defining a file view, and making a level-3 I/O
request for the distributed-array example of Figure~\ref{fig:access}.

\begin{figure}
\input{example.c}
\caption{Detailed code for the distributed-array example of
Figure~\protect\ref{fig:access} using a level-3 request}
\label{fig:example}
\end{figure}

\section{ROMIO Implementation of MPI-IO \label{sec:impl}}
We have developed a freely
available, portable implementation of MPI-IO, called
ROMIO~\cite{romio-www,thak99b}. It runs on at least the following
machines: IBM SP; Intel Paragon; Cray T3E; HP Exemplar; SGI
Origin2000; NEC SX-4; other symmetric multiprocessors from HP, SGI,
Sun, DEC, and IBM; and networks of workstations (Sun, SGI, HP, IBM,
DEC, Linux, and FreeBSD). Supported file systems are IBM PIOFS, Intel
PFS, HP HFS, SGI XFS, NEC SFS, NFS, PVFS, and any Unix file system
(UFS).

\begin{figure}
\centerline{\psfig{figure=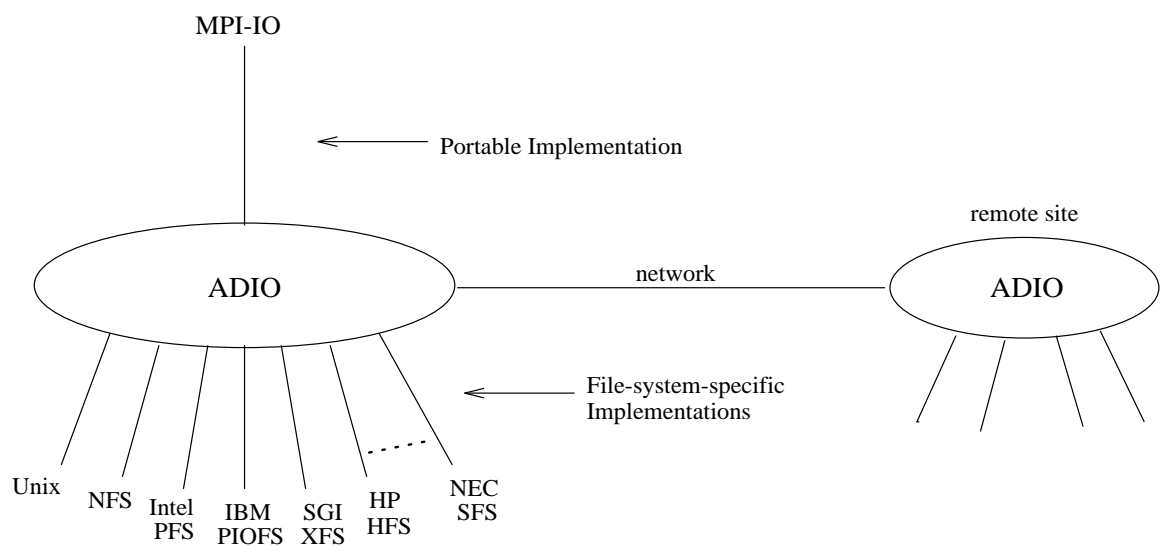,width=6in}}
\caption{ROMIO Architecture: MPI-IO is implemented portably on top of
an abstract-device interface called ADIO, and ADIO is optimized
separately for different file systems.}
\label{fig:adio}
\end{figure}

A key component of ROMIO that enables such a portable MPI-IO
implementation is an internal layer called ADIO~\cite{thak96e}. ADIO,
an abstract-device interface for I/O, is a mechanism for implementing
parallel I/O APIs portably on multiple file systems. ADIO
consists of a small set of basic functions for parallel I/O. We
have implemented MPI-IO portably on top of ADIO, and only ADIO is
implemented separately on each different file system (see
Figure~\ref{fig:adio}).  ADIO thus separates the machine-dependent and
machine-independent aspects involved in implementing MPI-IO.

The following two sections describe the optimizations ROMIO performs
for noncontiguous I/O requests, namely, data sieving for noncontiguous
requests from one process and collective I/O for noncontiguous
requests from multiple processes.

\section{Data Sieving \label{sec:siev}}

To reduce the effect of high I/O latency, it is critical to make as
few requests to the file system as possible. Data sieving is a
technique that enables an implementation to make a few large,
contiguous requests to the file system even if the user's request
consists of several small, noncontiguous accesses. Data sieving was
used in the PASSION I/O library~\cite{thak94f,thak96a} to access
sections of out-of-core arrays. We have extended it in ROMIO
to handle {\em any} general noncontiguous access pattern (as can be
described by an MPI datatype) and to use only a constant amount of
extra memory regardless of the access pattern. The user can control
the memory usage dynamically by setting a runtime parameter. 

\begin{figure}
\centerline{\psfig{figure=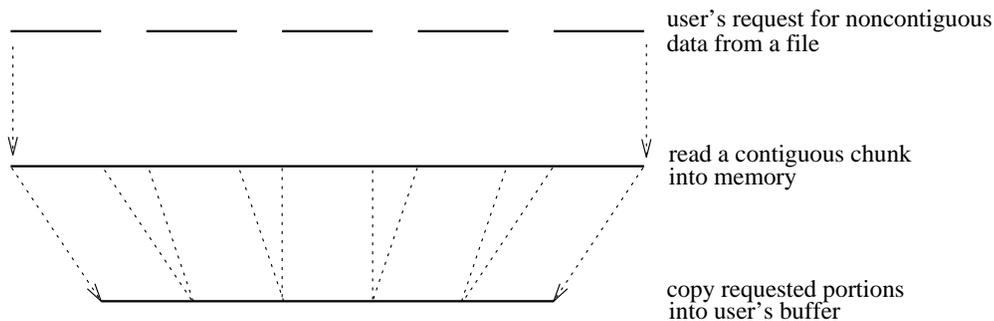,width=5in}}
\caption{Data sieving}
\label{fig:siev}
\end{figure}

Figure~\ref{fig:siev} illustrates the basic idea of data sieving.
Assume that the user has made a single read request for five
noncontiguous pieces of data. Instead of reading each noncontiguous
piece separately, ROMIO reads a single contiguous chunk
of data starting from the first requested byte up to the last
requested byte into a temporary buffer in memory. It then extracts the
requested portions from the temporary buffer and places them in the
user's buffer. The user's buffer happens to be contiguous in this
example, but it could well be noncontiguous.

A potential problem with this simple algorithm is its memory
requirement. The temporary buffer into which data is first read must
be as large as the {\em extent} of the user's request, where extent is
defined as the total number of bytes between the first and last byte
requested (including holes). The extent can potentially be very
large---much larger than the amount of memory available for the
temporary buffer---because the holes (unwanted data) between the
requested data segments could be very large. The basic algorithm,
therefore, must be modified to make its memory requirement independent
of the extent of the user's request.

ROMIO uses a user-controllable parameter that defines the maximum
amount of contiguous data that a process can read at a time during
data sieving. This value also represents the maximum size of the
temporary buffer. The default value is 4~Mbytes (on each process), but
the user can change it at run time via MPI-IO's hints mechanism. If
the extent of the user's request is larger than the value of this
parameter, ROMIO performs data sieving in parts, reading only as much
data at a time as defined by the parameter.

The advantage of data sieving is that data is always accessed in large
chunks, although at the cost of reading more data than needed. For
many common access patterns, the holes between useful data are not
unduly large, and the advantage of accessing large chunks far
outweighs the cost of reading extra data. In some access patterns,
however, the holes are so large that the cost of reading the
extra data outweighs the cost of accessing large chunks. The BTIO
benchmark (see Section~\ref{sec:perf}), for example, has such an
access pattern. An intelligent data-sieving algorithm can handle
such cases as well. The algorithm can analyze the user's request and
calculate the sizes of holes in it. Based on empirically determined
prior knowledge of how large holes can get before data sieving is no
longer beneficial, the algorithm can decide whether to perform data
sieving or access each contiguous data segment separately. 

Data sieving can similarly be used for writing data. A
read-modify-write must be performed, however, to avoid destroying the
data already present in the holes between contiguous data
segments. For writing with data sieving, ROMIO first reads a
contiguous chunk of data from the file into a temporary buffer in
memory, copies data from the user's buffer into appropriate locations
in the temporary buffer, and then writes the temporary buffer back to
the file. The portion of the file being accessed must also be locked
during the read-modify-write to prevent concurrent updates by other
processes.

ROMIO also uses another user-controllable parameter that defines the
maximum amount of contiguous data that a process can write at a time
during data sieving. This parameter, by default, has a smaller value
than the one used for reading, because writing involves locking the
region of the file being accessed. If the region being locked is too
large, many processes remain idle waiting for the lock to be
released. Consequently, parallelism in I/O is lost, and performance
decreases. On the other hand, if the region being locked is too small,
there is greater parallelism, but the size of each I/O access also
decreases, and performance is again adversely affected. In other
words, a compromise is needed between allowing greater concurrency and
having large access sizes. We determined experimentally that a write
size of 512~Kbytes provides a good trade-off between the two
conflicting goals and gives good performance. ROMIO therefore sets the
default value of the maximum buffer size for writing to
512~Kbytes. The user can, of course, change this value at run time.

One could argue that most file systems perform data sieving anyway
because they perform caching. That is, even if the user makes many
small I/O requests, the file system always reads multiples of disk
blocks and may also perform a read-ahead. The user's requests,
therefore, may be satisfied out of the file-system cache. Our
experience, however, has been that the cost of making many system
calls, each for small amounts of data, is extremely high, despite the
caching performed by the file system. In most cases, it is more
efficient to make a few system calls for large amounts of data and
extract the needed data. (See the performance results in
Section~\ref{sec:perf}.)

ROMIO performs data sieving when the user makes a level-2
(noncontiguous, noncollective) MPI-IO request. For level-3 requests
(noncontiguous, collective), data sieving is used within the
collective I/O implementation to perform the local I/O on each
process, as explained in the next section.

\section{Collective I/O \label{sec:coll}}
The preceding section explained how data sieving can be used to
optimize I/O when the entire (noncontiguous) access information of a
single process is known. Further optimization is possible if the
implementation is given the entire access information of a group of
processes. Such optimization is broadly referred to as collective I/O.

In many parallel applications, although each process may need to
access several noncontiguous portions of a file, the requests of
different processes are often interleaved and may together span large
contiguous portions of the file. If the user provides the MPI-IO
implementation with the entire access information of a group of
processes, the implementation can improve I/O performance
significantly by merging the requests of different processes and
servicing the merged request, that is, by performing collective I/O.

Collective I/O can be performed in different ways and has been studied
by many researchers in recent years. It can be done at the disk level
(disk-directed I/O~\cite{kotz97a}), at the server level
(server-directed I/O~\cite{seam96a,seam95b}), or at the client level
(two-phase I/O~\cite{rosa93a,thak96f} or collective
buffering~\cite{nitz97a}). Each method has its advantages and
disadvantages. Since ROMIO is a portable, user-level library with no
separate I/O servers, it performs collective I/O at the client level
by using a generalized version of two-phase I/O.

ROMIO performs collective I/O when the user makes level-3 MPI-IO
requests. Most level-1 requests do not contain enough information for
ROMIO to perform collective optimizations, and ROMIO therefore implements them
internally as level-0 requests. Some level-1 requests, such as those
that represent a read-broadcast type of access pattern, are optimized
collectively, however.

\subsection{Two-Phase I/O}
\begin{figure}
\centerline{\psfig{figure=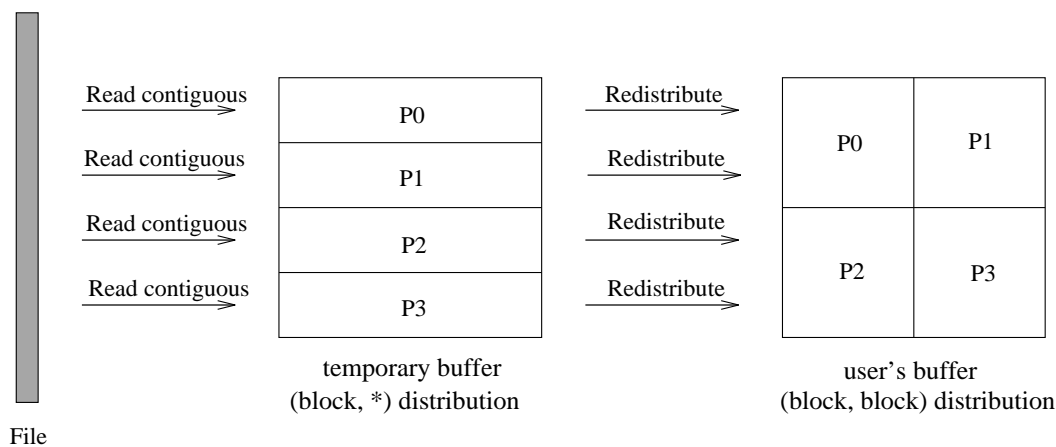,width=5.5in}}
\caption{Reading a distributed array by using two-phase I/O}
\label{fig:twophase}
\end{figure}

Two-phase I/O was first proposed in~\cite{rosa93a} in the context of
accessing distributed arrays from files. Consider the example of
reading a two-dimensional array from a file into a (block,block)
distribution in memory, as shown in Figure~\ref{fig:twophase}. Assume
that the array is stored in the file in row-major order. As a result
of the distribution in memory and the storage order in the file, the
local array of each process is located noncontiguously in the file:
each row of the local array of a process is separated by rows from the
local arrays of other processes. If each process tries to read each
row of its local array individually, the performance will be poor
because of the large number of relatively small I/O requests. Note,
however, that all processes together need to read the entire file, and
two-phase I/O uses this fact to improve performance as explained below.

If the entire I/O access pattern of all processes is known to the
implementation, the data can be accessed efficiently by splitting the
access into two phases. In the first phase, processes access data
assuming a distribution in memory that results in each process making
a single, large, contiguous access. In this example, such a
distribution is a row-block or (block,*) distribution. In the second
phase, processes redistribute data among themselves to the desired
distribution. The advantage of this method is that by making all file
accesses large and contiguous, the I/O time is reduced significantly.
The added cost of interprocess communication for redistribution is
(almost always) small compared with the savings in I/O time.

The basic two-phase method was extended in~\cite{thak96f} to access
sections of out-of-core arrays. Since MPI-IO is a general parallel I/O
interface, I/O requests in MPI-IO can represent {\em any} access
pattern, not just sections of arrays. The two-phase method
in~\cite{thak96f} must therefore be generalized to handle any
noncontiguous I/O request. We have implemented such a scheme in ROMIO.

Two-phase I/O does increase the memory requirements of a program. For
reading a distributed array, for example, the amount of extra memory
needed on each process (to store the data read in the first phase) is
equal to the size of the local array itself. Since this amount of
memory may not be available, the basic two-phase algorithm must be
modified to read and communicate smaller parts of the array at a
time. Similarly, on machines in which the I/O performance does not
scale with the number of processes making simultaneous file accesses,
it may be beneficial to have only a subset of processes perform
I/O, with the remaining processes participating only in the
redistribution phase. All these generalizations---any access pattern,
fixed memory requirement, and variable number of processes performing
I/O---are incorporated in ROMIO's collective I/O implementation.

\subsection{Generalized Two-Phase I/O in ROMIO}
ROMIO uses two user-controllable parameters for collective I/O: the
number of processes that should directly access the file and the
maximum size on each process of the temporary buffer needed for
two-phase I/O. By default, all processes perform I/O in the I/O phase,
and the maximum buffer size is 4~Mbytes per process. The user can
change these values at run time via MPI-IO's hints mechanism (see
Figure~\ref{fig:info}). 

\begin{figure}[t]
\input{info.c}
\caption{Example showing how to specify hints in MPI-IO. The two
collective I/O hints, {\tt cb\_buffer\_size} and {\tt cb\_nodes}, are
predefined hints in MPI-IO; the two data-sieving hints, {\tt
ind\_rd\_buffer\_size} and {\tt ind\_wr\_buffer\_size}, are additional
hints that ROMIO supports.}
\label{fig:info}
\end{figure}

\begin{figure}[t]
\centerline{\psfig{figure=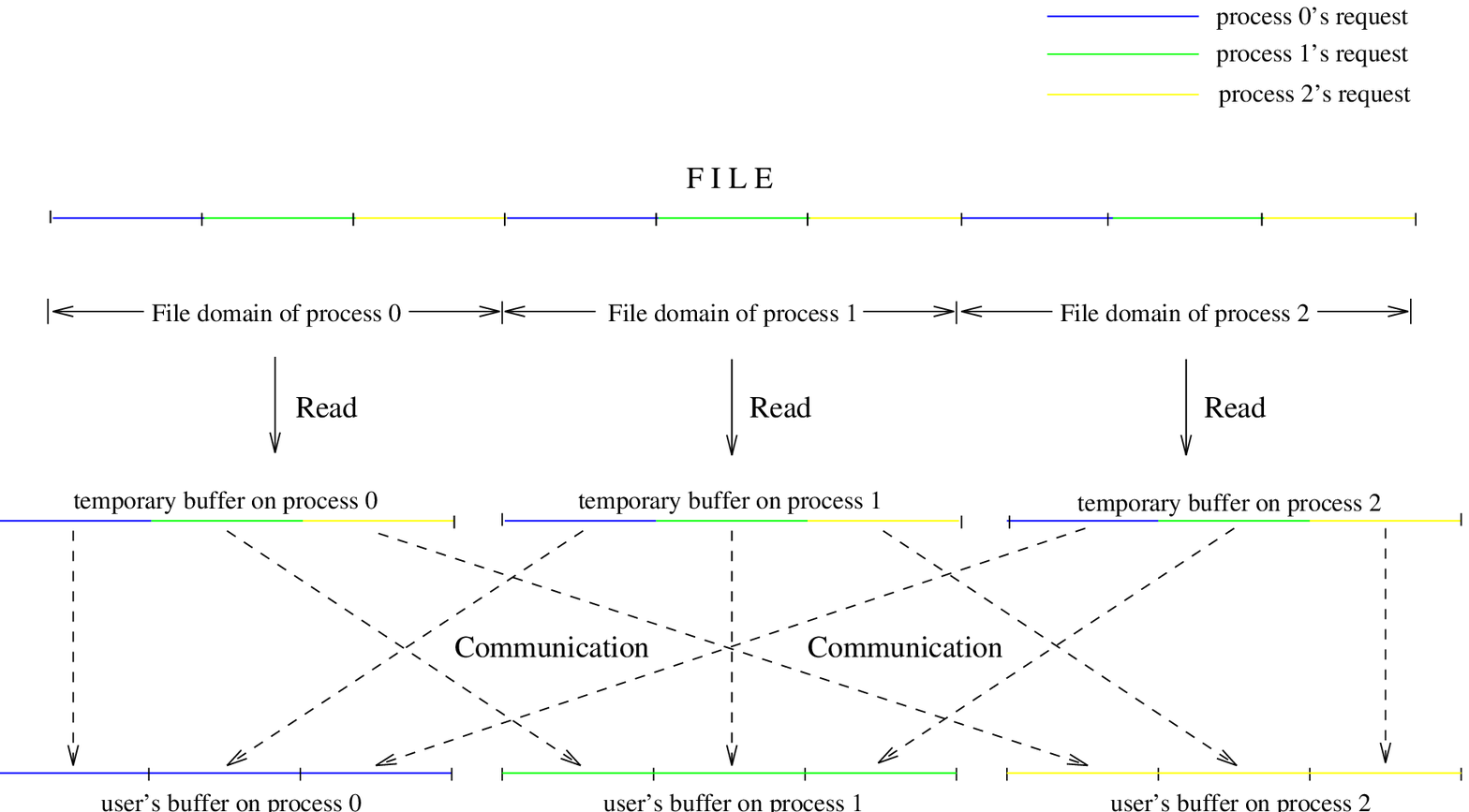,height=4in,width=6.5in}}
\caption{A simple example illustrating how ROMIO performs a collective read}
\label{fig:romiocoll}
\end{figure}

We first explain the algorithm that ROMIO uses for collective reads
and then describe how the algorithm differs for collective writes.
Figure~\ref{fig:romiocoll} shows a simple example that illustrates how
ROMIO performs a collective read. In this example, all processes
perform I/O, and each process is assumed to have as much memory as
needed for the temporary buffer.

In MPI-IO, the collective I/O function called by a process specifies
the access information of that process only. If an MPI-IO
implementation needs the access information of all processes
participating in a collective I/O operation, it must gather the
information from those processes during the execution of the
collective I/O function. Also, file accesses in collective I/O refer
to accesses from multiple processes to a {\em common} file.

\subsubsection{Collective Reads}
In ROMIO's implementation of collective reads, each process first
analyzes its own I/O request and creates a list of offsets and a list
of lengths, where {\tt length[i]} gives the number of bytes that the
process needs from location {\tt offset[i]} in the file. Each process
also calculates the locations of the first byte (start offset) and the
last byte (end offset) it needs from the file and then broadcasts
these two offsets to other processes. As a result, each process has
the start and end offsets of all processes.

In the next step, each process tries to determine whether this
particular access pattern can benefit from collective I/O, that is,
whether the accesses of any of the processes are interleaved in the
file.  Since an exhaustive check can be expensive, each process checks
only whether, for any two processes with consecutive ranks ($i$ and
$i+1$), the following expression is true: (start-offset$_{i+1}$ $<$
end-offset$_i$). If the expression is not true, each process concludes
that collective I/O will not improve performance for this particular
access pattern, since the requests of different processes cannot be
merged. In such cases, each process just calls the corresponding
independent I/O function, which uses data sieving to optimize
noncontiguous requests.

If the above expression is true, the processes perform collective I/O
as follows.  Portions of the file are ``assigned'' to each process
such that, in the I/O phase of the two-phase operation, a process will
access data only from the portion of the file assigned to it. This
portion of the file assigned to a process is called the process's {\em
file domain}. If a process needs data located in another process's
file domain, it will receive the data from the other process during
the communication phase of the two-phase operation. Similarly, if this
process's file domain contains data needed by other processes, it must
send this data to those processes during the communication phase.

File domains are assigned as follows. Each process calculates the
minimum of the start offsets and the maximum of the end offsets of all
processes. The difference between these two offsets gives the total
extent of the combined request of all processes. The file domain of
each process is obtained by dividing this extent equally among the
processes. For example, if the combined request of all processes spans
from offset~100 to offset~399 in the file, and there are three
processes, the file domain of process~0 will be from offset~100
to~199; the file domain of process~1 will be from offset~200 to~299;
and the file domain of process~2 will be from offset~300 to~399.

When file domains are selected in this manner, the file domain of a
process may not contain data needed by any process (e.g., if the
access pattern has large holes). In such a case, the process will not
perform any I/O and will participate only in communication. (It is
possible to design a more intelligent file-domain selection scheme
that analyzes the access pattern and then assigns file domains in a
manner that ensures an even balance of the I/O workload and/or reduces
the communication needs.)

After the file domains are determined, each process calculates in
which other process's file domain its own I/O request (or a portion of
it) is located. For each such process, it creates a data structure
containing a list of offsets and lengths that specify the data needed
from the file domain of that process. It then sends this access
information to the processes from which it expects to receive
data. Similarly, other processes that need data from the file domain
of this process send the corresponding access information to this
process. After this exchange has taken place, each process knows what
portions of its file domain are needed by other processes and by
itself. It also knows which other processes are going to send the data
that it needs.

The next step is to read and communicate the data. This step consumes
the majority of the time because all the I/O and data communication
takes place here.  Note that the communication in earlier steps
involved only access information. The access information is usually
much smaller than actual data, unless the access pattern is so
irregular that an index is needed to represent the location of every
basic datatype needed from the file.

As mentioned above, ROMIO performs the read-and-communicate step in
several parts to reduce its memory requirement. Each process first
calculates the offsets corresponding to the first and last bytes
needed (by any process) from its file domain. It then divides the
difference between these offsets by the maximum size allowed for the
temporary buffer (4~Mbytes by default). The result is the number of
times ({\tt ntimes}) it needs to perform I/O. All processes then
perform a global-maximum operation on {\tt ntimes} to determine the
maximum number of times ({\tt max\_ntimes}) any process needs to
perform I/O. Even if a process has completed all the I/O needed from
its own file domain, it may need to participate in communication
operations thereafter to receive data from other processes. Each
process must therefore be ready to participate in the communication
phase {\tt max\_ntimes} number of times.

For each of the {\tt ntimes} I/O operations, a process does the
following operations. It checks whether the current portion of its
file domain (no larger than the maximum buffer size) has data that any
process needs, including itself. If it does not have such data, the
process does not need to perform I/O in this step; it then checks
whether it needs to receive data from other processes, as explained
below.  If it does have such data, it reads with a single I/O function
call all the data from the first offset to the last offset needed from
this portion of the file domain into a temporary buffer in memory. The
process effectively performs data sieving, as the data read may
include some unwanted data. Now the process must send portions of the
data read to processes that need them.

Each process first informs other processes how much data it is going
to send to each of them. The processes then exchange data by first
posting all the receives as nonblocking operations, then posting all
the nonblocking sends, and finally waiting for all the nonblocking
communication to complete.  MPI derived datatypes are used to send
noncontiguous data directly from the temporary buffer to the
destination process. On the receive side, if the user has asked for
data to be placed contiguously in the user-supplied buffer, the data
is received directly into the user's buffer. If data is to be placed
noncontiguously, the process first receives data into a temporary
buffer and then copies it into the user's buffer. (Since data is
received in parts over multiple communication operations from
different processes, we found this approach easier than creating
derived datatypes on the receive side.)

Each process performs I/O and communication {\tt ntimes} number of
times and then participates only in the communication phase for the
remaining {\tt (max\_ntimes - ntimes)} number of times. In some of
these remaining communication steps, a process may not receive any
data; nevertheless, the process must check whether it is going to
receive data in a particular step.

\subsubsection{Collective Writes}
The algorithm for collective writes is similar to the one for
collective reads except that the first phase of the two-phase
operation is communication and the second phase is I/O. In the I/O
phase, each process checks to see whether any holes (gaps) exist in
the data it needs to write. If holes exist, it performs a
read-modify-write; otherwise it performs only a write. During the
read-modify-write, a process need not lock the region of the file
being accessed (unlike in independent I/O), because the process is
assured that no other process involved in the collective I/O operation
will directly try to access the data located in this process's file
domain. The process is also assured that concurrent writes from
processes other than those involved in this collective I/O operation
will not occur, because MPI-IO's consistency semantics~\cite{mpi97a}
do not automatically guarantee consistency for such writes. (In such
cases, users must use {\tt MPI\_File\_sync} and ensure that the
operations are not concurrent.)

\subsubsection{Performance Issues}
Even if I/O is performed in large contiguous chunks, the performance
of the collective I/O implementation can be significantly affected by
the amount of buffer copying and communication. We were able to
improve ROMIO's collective I/O performance by as much as 50\% on some
machines by tuning the implementation to minimize buffer copying,
minimize the number of communication calls, and use the right set of
MPI communication primitives.

Initially, in each of the communication steps, we always received data
into a temporary buffer and then copied it into the user's buffer. We
realized later that this copy is needed only when the user's buffer is
to be filled noncontiguously. In the contiguous case, data can be
received directly into the appropriate location in the user's
buffer. We similarly experimented with different ways of communicating
data in MPI and measured the effect on overall collective I/O
performance with different MPI implementations and on different
machines. We selected nonblocking communication with the receives
posted first and then the sends, a strategy that performs well on most
systems. It may be possible, however, to tune the communication
further on some machines by posting the sends before the receives or
by using MPI's persistent requests.

\subsubsection{Portability Issues}
We were able to implement these optimizations portably, and without
sacrificing performance, by using ADIO as a portability layer for I/O
(see Section~\ref{sec:impl}) and by using MPI for communication.
Data sieving and collective I/O are implemented as ADIO
functions~\cite{thak96e}; data sieving is used in the ADIO functions
that read/write noncontiguous data, and collective I/O is used in
ADIO's collective I/O functions. Both these optimizations ultimately
make contiguous I/O requests to the underlying file system, which are
implemented by using ADIO's contiguous I/O functions. The
contiguous I/O functions, in turn, are implemented by using the
appropriate file-system call for each different file system.

\section{Performance Evaluation \label{sec:perf}}
We describe the three applications used in the performance
experiments, the machines on which we ran the applications, and the
set of experiments performed. We then present and analyze the
performance results.

\subsection{Applications}
The first application we used is DIST3D, a template representing the
I/O access pattern in an astrophysics application,
ASTRO3D~\cite{thak96c}, from the University of Chicago. It measures
the performance of reading/writing a three-dimensional array
distributed in a (block,block,block) fashion among processes from/to a
file containing the global array in row-major order.

The second application is the BTIO benchmark~\cite{fine96a} from NASA
Ames Research Center, which simulates the I/O required by a
time-stepping flow solver that periodically writes its solution
matrix. The solution matrix is distributed among processes by using a
multipartition distribution~\cite{brun88a} in which each process is
responsible for several disjoint subblocks of points (cells) of the
grid. The cells are arranged such that, for each direction of the
solve phase, the cells belonging to a certain process will be evenly
distributed along the direction of solution.  The solution matrix is
stored on each process as $C$ three-dimensional arrays, where $C$ is
the number of cells on each process. (The arrays are actually
four dimensional, but the first dimension has only five elements and
is not distributed.) Data is stored in the file in an order
corresponding to a column-major ordering of the global solution
matrix. Note that this distribution is different from the
(block,block,block) distribution of DIST3D.  The benchmark performs
only writes, but we modified it to perform reads also, in order to
measure the read bandwidths.

The third application we used is an unstructured code (which we call
UNSTRUC) written by Larry Schoof and Wilbur Johnson of Sandia National
Laboratories. It is a synthetic benchmark that emulates the I/O access
pattern in unstructured-grid applications. It generates a random
irregular mapping from the local one-dimensional array of a process to
a global array in a common file shared by all processes. The mapping
specifies where each element of the local array is located in the
global array. The size of each element can also be varied in the
program.

\subsection{Machines}
We ran the codes portably and measured the performance on five
different parallel machines: the HP Exemplar and SGI Origin2000 at the
National Center for Supercomputing Applications (NCSA), the IBM SP at
Argonne National Laboratory, the Intel Paragon at the California Institute
of Technology, and the NEC SX-4 at the National Aerospace Laboratory
(NLR) in Holland. These machines cover almost the entire spectrum of
high-performance systems, and they represent
distributed-memory, shared-memory, and parallel vector architectures.
They also represent a wide variation in I/O architecture, from the
``traditional'' parallel file systems on distributed-memory machines
such as the SP and Paragon, to the so-called high-performance file
systems on shared-memory machines such as the Origin2000, Exemplar,
and SX-4.

We used the native file systems on each machine: HFS on the Exemplar,
XFS on the Origin2000, PIOFS on the SP, PFS on the Paragon, and SFS on
the SX-4. At the time we performed the experiments, these file systems
were configured as follows: HFS on the Exemplar was configured on
twelve disks; XFS on the Origin2000 had two RAID units with SCSI-2
interfaces; the SP had four servers for PIOFS, each server with four
SSA disks attached to it in one SSA loop; the Paragon had 64~I/O nodes
for PFS, each with an individual Seagate disk; and SFS on the NEC SX-4
was configured on a single RAID unit comprising sixteen SCSI-2 data
disks.

\subsection{Experiments}
We modified the I/O portions of these applications to correspond to
each of the four levels of requests (see Section~\ref{sec:class}) and
ran the programs on all five machines. In all experiments, we used the
default values of the sizes of the internal buffers ROMIO uses for
data sieving and collective I/O (see Sections~\ref{sec:siev}
and~\ref{sec:coll}). We also used the default values of the
file-striping parameters on all file systems. On PFS and PIOFS the
default striping unit was 64~Kbytes.

% The space of experimentation is
% very large, with many parameters that can be varied. While it would be
% interesting to see the effect of scaling the problem size and the
% number of processors, we chose to limit the experiments to large
% problem sizes on a large number of processors in order to limit the
% number of results to be gathered and presented. These experiments also
% give us an idea of the maximum I/O bandwidth that can realistically be
% achieved on these machines---a fact we were interested in knowing.

% We measured the write performance without explicitly calling an {\tt
% MPI\_File\_sync} to flush all cached data to disk. Some of the
% performance results may therefore include caching performed by the
% file system. We did not include {\tt MPI\_File\_sync} in the
% measurements because users most often do not perform a file sync;
% they just open, read/write, and close. 

On each machine, we used as many processors as we could reasonably
access. We also tried to use the same number of processors on a given
machine for each application but were at times constrained by the
application's requirements: BTIO requires that the number of
processors be a perfect square, whereas UNSTRUC requires that the
number of processors be a power of two. On some machines,
therefore, we could not use the same number of processors for both
BTIO and UNSTRUC; for example, on the NEC SX-4 we had to run BTIO on
9~processors and UNSTRUC on 8~processors. 

The access patterns in DIST3D and BTIO are such that level-1 requests
cannot be optimized with collective I/O. In such cases, ROMIO
internally translates level-1 requests into level-0 requests (with
some overhead incurred in analyzing the level-1 request).  In UNSTRUC,
the I/O access pattern is irregular, and the granularity of each
access is very small (64~bytes). Level-0/1 requests are not feasible
for this kind of application because they take an excessive amount of
time. Therefore, we present results with level-0, level-2, and level-3
requests for DIST3D and BTIO, and with level-2 and level-3 requests for
UNSTRUC.

\subsection{Results}
\begin{figure}[t]
\hskip -0.05in
\begin{minipage}[b]{0.5\linewidth}
\centerline{\psfig{figure=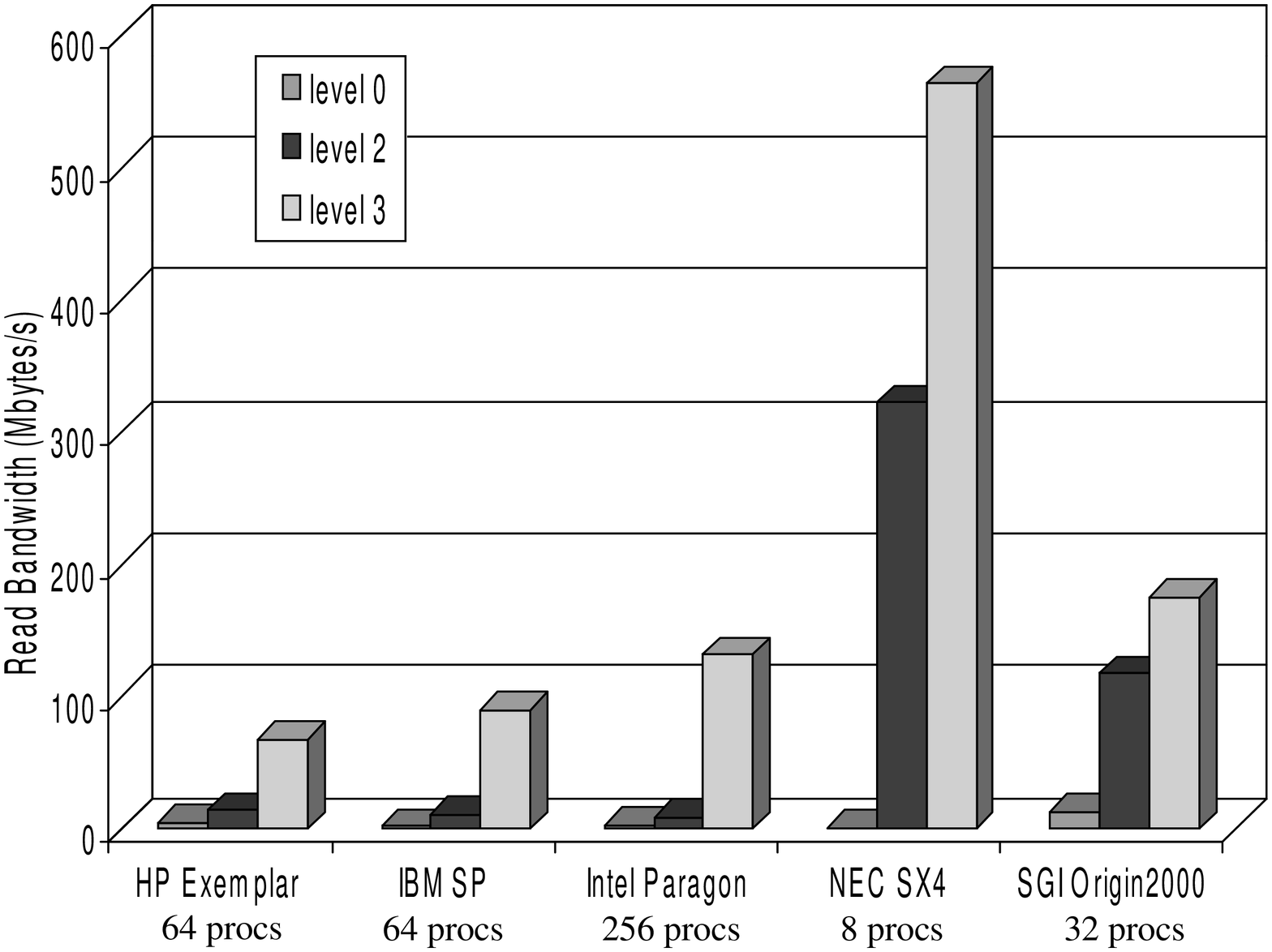,height=2.7in,width=3.4in}}
\end{minipage}
%\hskip 0.05in
\begin{minipage}[b]{0.5\linewidth}
\centerline{\psfig{figure=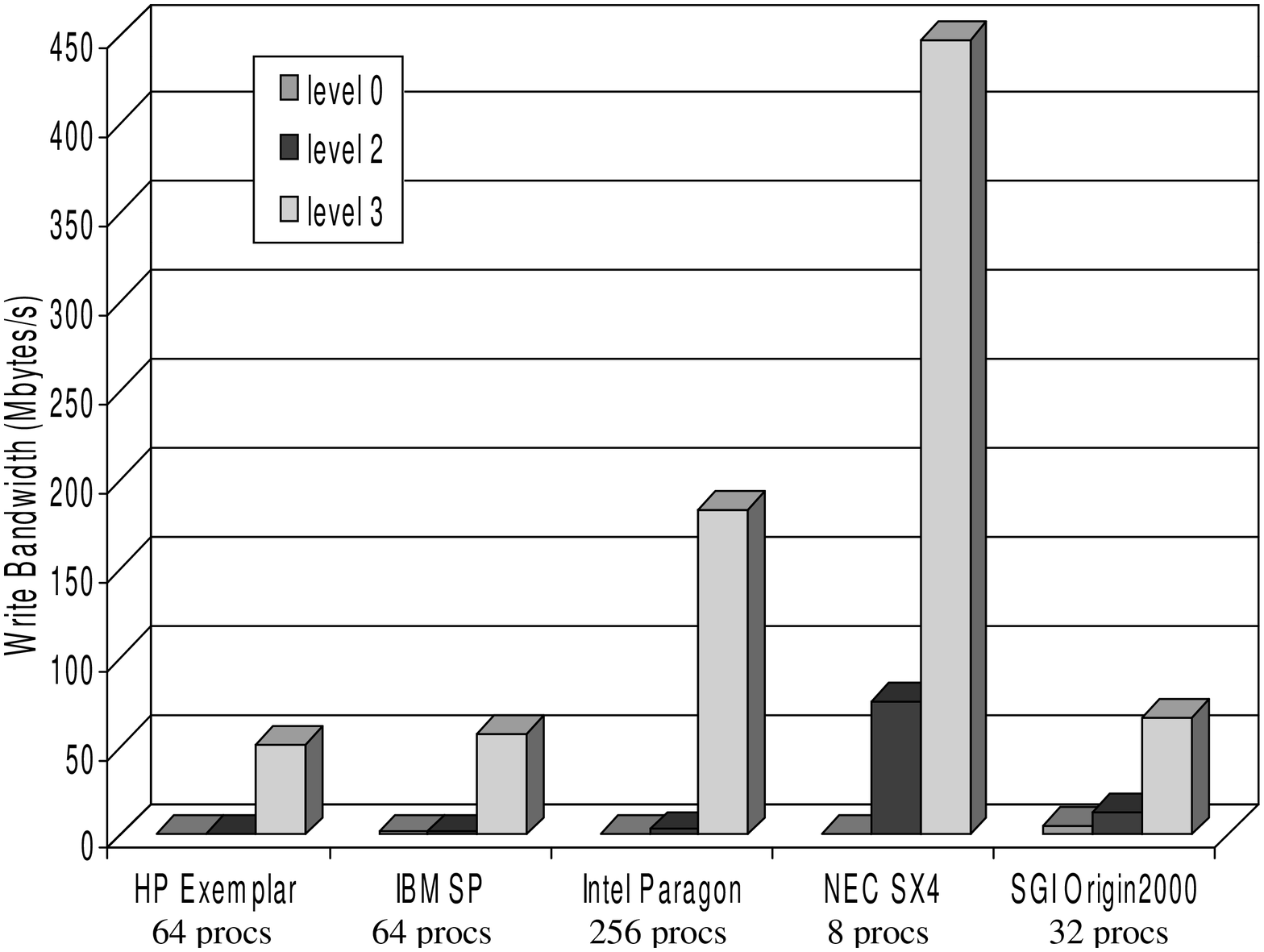,height=2.7in,width=3.4in}}
\end{minipage}
\caption{Performance of DIST3D (array size 512x512x512 integers =
512~Mbytes). Level-1 results are not shown because, for this access
pattern, ROMIO simply translates level-1 requests internally into
level-0 requests.} 
\label{fig:array}
\end{figure}

Figure~\ref{fig:array} shows the read and write bandwidths for
DIST3D. We calculated the bandwidth as the total data transferred by all
processes divided by the maximum of the time taken for I/O by any one process.
The performance with level-0 requests was, in general, very
poor because level-0 requests result in too many small read/write
calls. For level-2 requests---for which ROMIO performs data
sieving---the read bandwidth improved over level-0 requests by a
factor ranging from 2.6 on the HP Exemplar to 453 on the NEC
SX-4. Similarly, the write bandwidth improved by a factor ranging from
2.3 on the HP Exemplar to 121 on the NEC SX-4. On the IBM SP, however,
level-2 write requests performed the same as level-0 requests. This is
because ROMIO cannot perform data sieving for writing on the SP's
PIOFS file system, since PIOFS does not support file locking. On
PIOFS, ROMIO internally translates level-2 requests into level-0
requests.

The performance improved considerably with level-3 requests because 
ROMIO performs collective I/O in this case.
The read bandwidth improved by a factor of as much as
793 over level-0 requests (NEC SX-4) and as much as 14
over level-2 requests (Intel Paragon). Similarly, with level-3
requests, the write performance improved by a
factor of as much as 721 over level-0 requests (NEC
SX-4) and as much as 40 over level-2 requests (HP Exemplar).

\begin{figure}[t]
\hskip -0.05in
\begin{minipage}[b]{0.5\linewidth}
\centerline{\psfig{figure=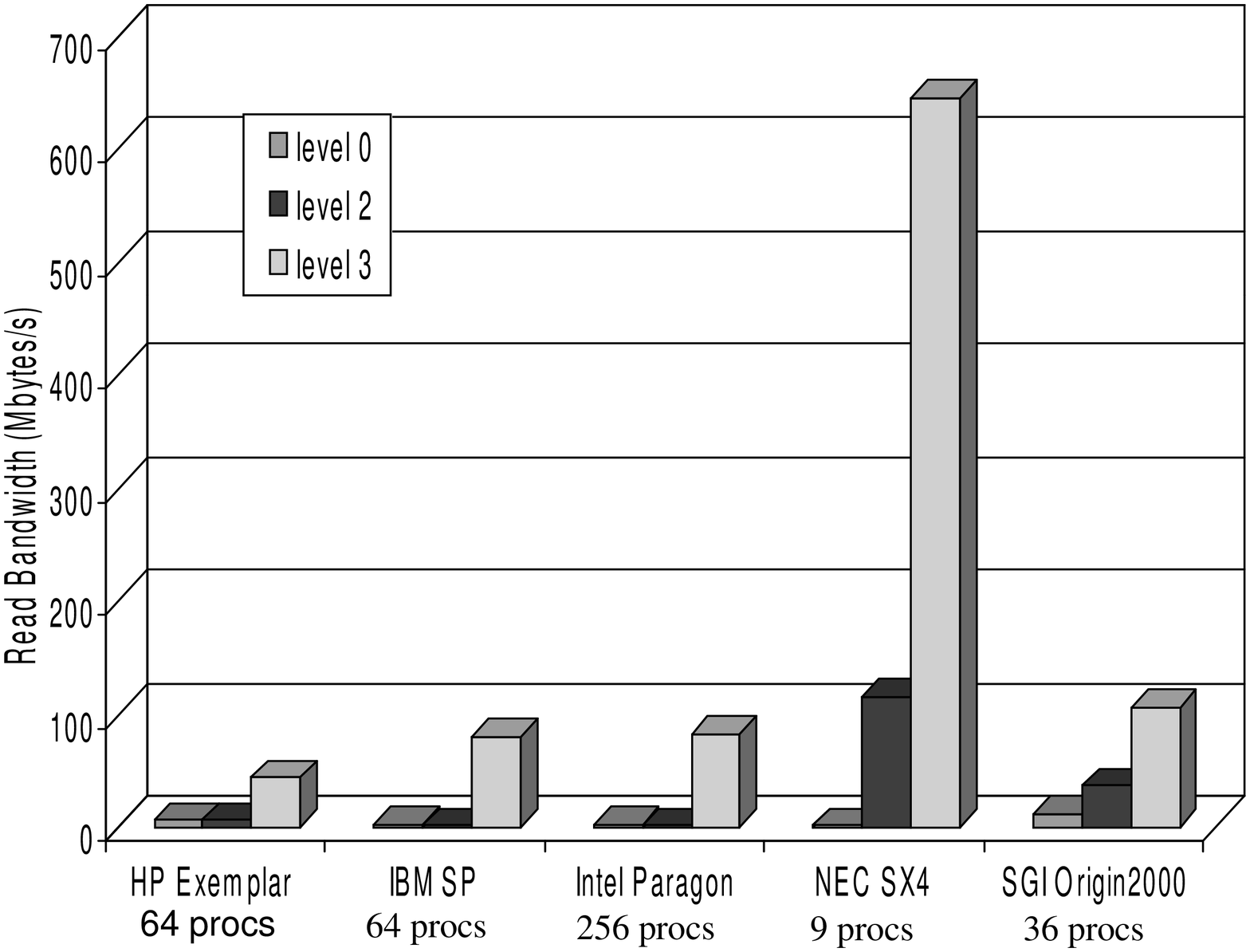,height=2.7in,width=3.4in}}
\end{minipage}
%\hskip 0.05in
\begin{minipage}[b]{0.5\linewidth}
\centerline{\psfig{figure=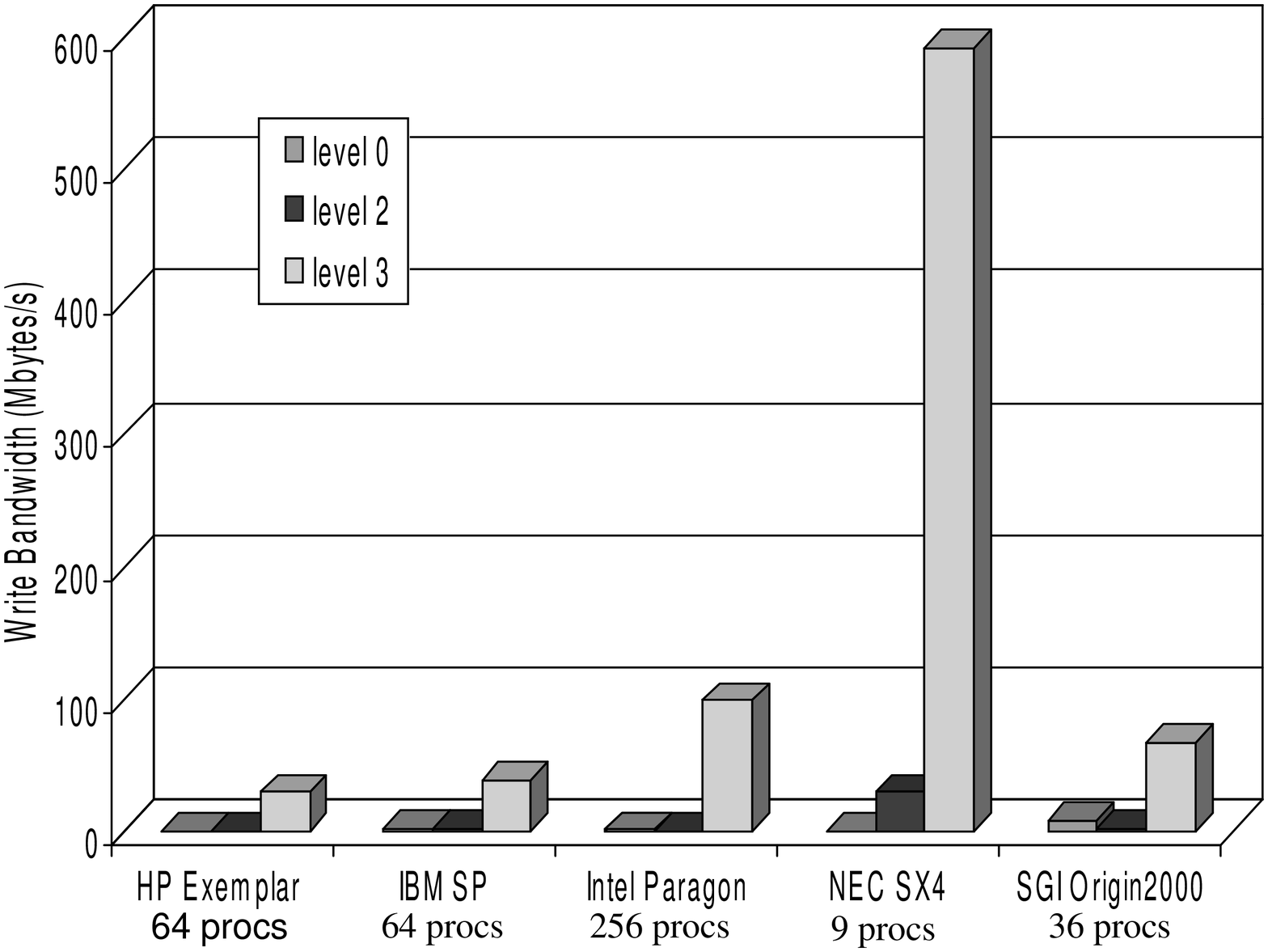,height=2.7in,width=3.4in}}
\end{minipage}
\caption{Performance of BTIO (Class C, problem size
5x162x162x162 double precision $\approx$ 162~Mbytes). Level-1 results
are not shown because, for this access pattern, ROMIO simply translates
level-1 requests internally into level-0 requests.}
\label{fig:btio}
\end{figure}

Figure~\ref{fig:btio} presents results
for Class C of the BTIO benchmark. For BTIO, level-0 requests
performed better than level-2 requests on three out of the five
machines. The reason is that the holes between data segments needed by
a process are large in BTIO---more than five times the size of the
data segment. As a result, a lot of unwanted data was accessed during
data sieving (level~2), resulting in lower performance than with Unix-style
accesses (level~0). As mentioned in Section~\ref{sec:siev}, an intelligent
data-sieving algorithm could detect such large holes and internally
perform Unix-style accesses. ROMIO's data-sieving algorithm does not
currently do this, however.

Level-3 requests  performed extremely well on BTIO
because no unwanted data was accessed during collective I/O and all
accesses were large. The performance improved by a factor of as much
as 512 over level-0 requests for reading and 597 for writing, both on the
NEC SX-4.

\begin{figure}[t]
\hskip -0.05in
\begin{minipage}[b]{0.5\linewidth}
\centerline{\psfig{figure=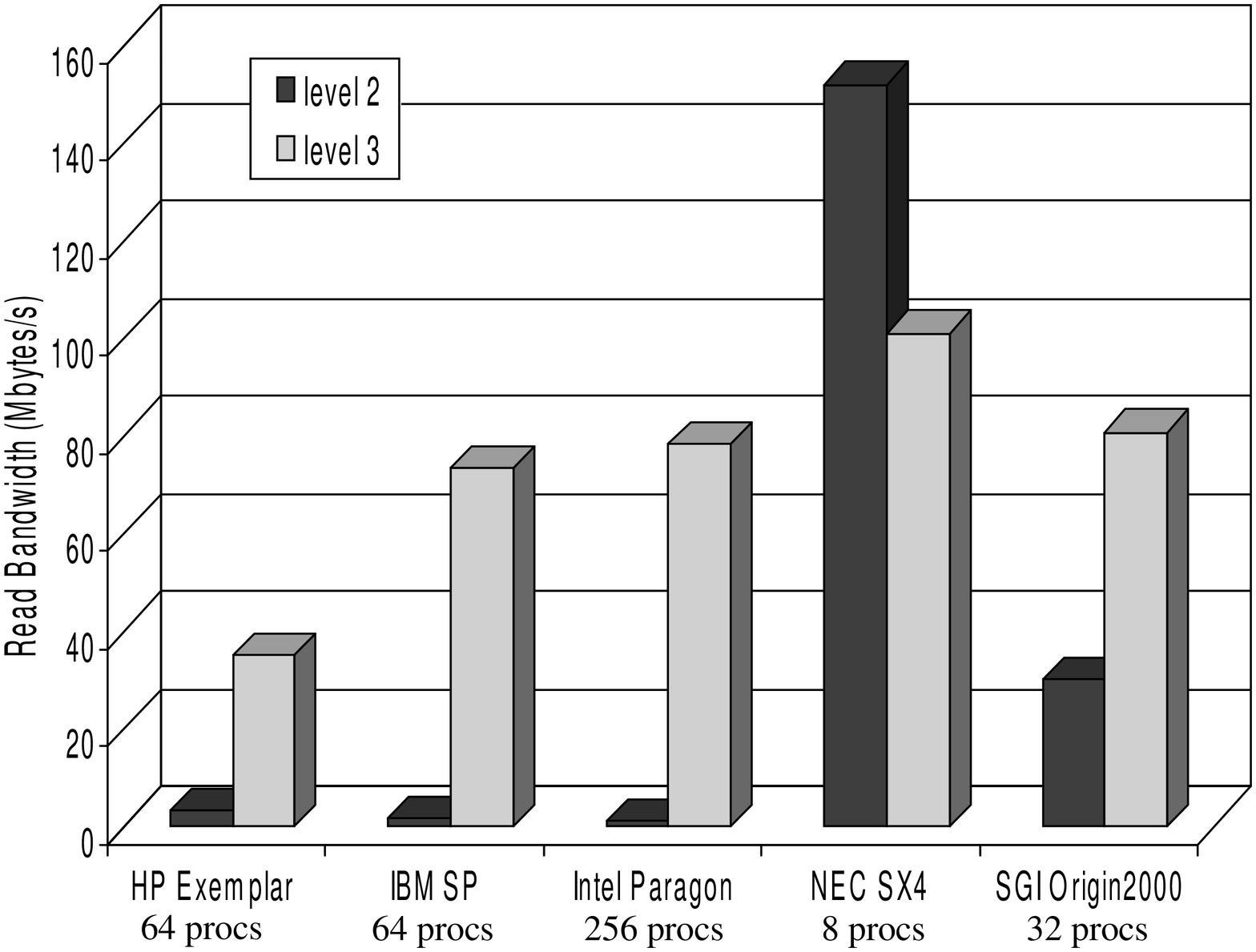,height=2.7in,width=3.4in}}
\end{minipage}
%\hskip 0.05in
\begin{minipage}[b]{0.5\linewidth}
\centerline{\psfig{figure=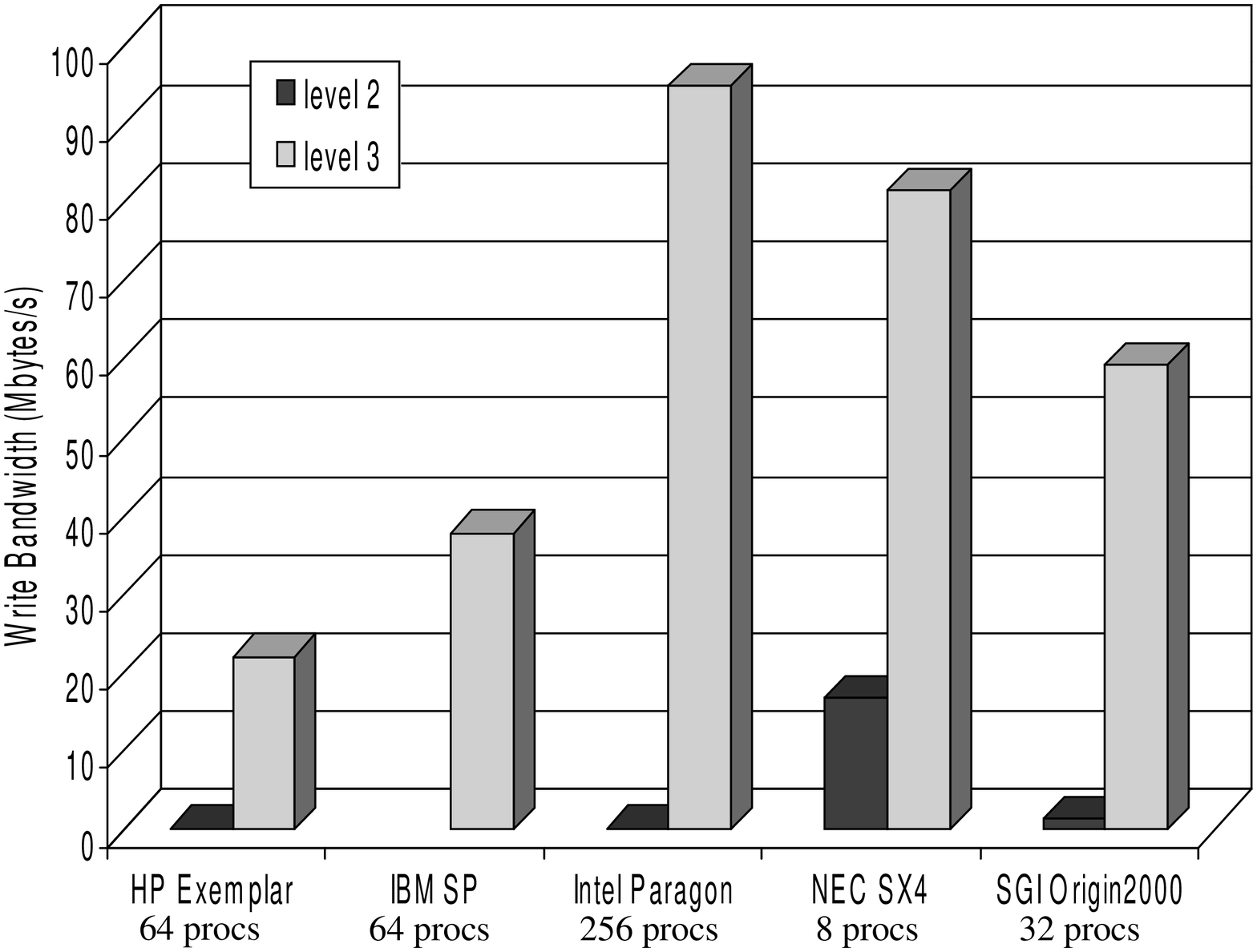,height=2.7in,width=3.4in}}
\end{minipage}
\caption{Performance of UNSTRUC. Level 0/1 results are not feasible
for this application because they take an excessive amount of time
because of the small granularity of each request. On the IBM SP,
because of the absence of file locking in the PIOFS file system, ROMIO
translates level-2 writes into level-0 writes, which are very slow in
this case. Hence, results for level-2 writes on the SP are not shown.}
\label{fig:irreg}
\end{figure}

Figures~\ref{fig:irreg} shows the read and write
bandwidths for UNSTRUC. We ran a problem size of 8 million
grid points on all machines except the Origin2000 where, because of memory
limitations imposed by the scheduler, we had to run a smaller problem
size of 4 million grid points.  Level-3 requests again performed much
better than level-2 requests, the only exception being for reads on
the NEC SX-4. In this case, because of the high read bandwidth of
NEC's Supercomputing File System (SFS), data sieving by itself
outperformed the extra communication required for collective I/O.

\subsection{Impact of Architecture and System Configuration}
The above results show that although level 3 performed the best on
each machine, there was a wide variation in the performance of the
applications among the different machines. This variation is because
the machines were configured with different amounts and types of I/O
hardware (disks), different amounts of memory, and, of course, they
had different I/O architectures and file systems. Our goal in this
study was to compare the performance of the different levels of
requests on a given machine, rather than comparing the performance of
different machines. In general, the parallel I/O performance of a machine
depends on the following factors: 
\begin{itemize}
\item the I/O architecture;
\item the speed and amount of I/O hardware (disks, etc.);
\item how well the file system can handle concurrent reads and writes;
and
\item how well the file system's caching policies (read-ahead,
write-behind) work for the given application.
\end{itemize}
The performance on the NEC SX-4 was the best among the five
machines. We believe that is because the machine has high memory and
I/O bandwidth and it was configured with sufficient I/O hardware for
high performance. We believe the performance would have been similar
even if the machine had more processors.

\section{Conclusions \label{sec:conc}}
The results in the preceding section demonstrate that MPI-IO can
deliver good I/O performance to applications. To achieve high
performance with MPI-IO, however, users must use some of MPI-IO's
advanced features, particularly noncontiguous accesses and collective
I/O. By making level-3 MPI-IO requests (noncontiguous, collective), we
achieved I/O bandwidths on the order of hundreds of Mbytes/sec,
whereas with level-0 requests (Unix style) we achieved less than
15~Mbytes/sec even when using high-performance file systems.  With
level-3 requests, the bandwidth achieved was limited only by the I/O
capabilities of the machine and underlying file system.  We believe
that such performance improvements with level-3 requests can also be
expected in applications other than those considered in this paper. 

We have described in detail the optimizations ROMIO performs for
noncontiguous requests: data sieving and collective I/O. We note that,
to achieve high performance, these optimizations must be carefully
implemented to minimize the overhead of buffer copying and
interprocess communication. Otherwise, these overheads can impact
performance significantly.

To carry out these optimizations, an MPI-IO implementation needs some
amount of temporary buffer space, which reduces the total amount of
memory available to the application. The optimizations, however, can
be performed with a constant amount of buffer space that does not
increase with the size of the user's request. Our results demonstrate
that by allowing the MPI-IO implementation to use as little as
4~Mbytes of buffer space per process, which is a small amount on
today's high-performance machines, users can gain orders of magnitude
improvement in I/O performance.

We note that the MPI-IO standard does not {\em require} an
implementation to perform any of these optimizations. Nevertheless,
even if an implementation does not perform any optimization and
instead translates level-3 requests into several level-0 requests to
the file system, the performance would be no worse than if the user
made level-0 requests. Therefore, there is no reason not to
use level-3 requests (or level-2 requests where level-3 requests are
not possible).

\bibliographystyle{plain}
%\bibliography{/homes/thakur/tex/bib/papers}
\bibliography{references.bib}

\end{document}